\documentstyle[12pt]{article}
\newcommand{\be}{\begin{equation}} 
\newcommand{\nn}{\nonumber}
\newcommand{\bea}{\begin{eqnarray}}
\newcommand{\eea}{\end{eqnarray}}
\newcommand{\ba}{\begin{array}}
\newcommand{\ea}{\end{array}}
\newcommand{\ee}{\end{equation}}

\expandafter\ifx\csname mathbbm\endcsname\relax

\else

\fi \textheight 22cm \textwidth 15cm \topmargin 1mm \oddsidemargin
5mm \evensidemargin 5mm

\newcommand{\beas}{\begin{eqnarray*}}
\newcommand{\eeas}{\end{eqnarray*}}
\newcommand{\bes}{\begin{equation*}}
\newcommand{\ees}{\end{equation*}}

\newcommand{\lf}{\left}
\newcommand{\ri}{\right}
\newcommand{\f}{\frac}

\def\cO{{\cal O}}

\def\tr           {\mbox{\rm tr}\,}
\def\det           {\mbox{\rm det}\,}

\def\i2           {\mbox{$\frac{i}{2}$}}

\def\al           {\alpha}

\def\del           {\delta}

\def\et           {\eta}

\def\ga           {\gamma}

\def\la           {\lambda}
\def\lab          {\bar \la}

\def\ph           {\phi}

\def\sm        {\sqrt{\mb M}}

\def\pl           {\partial}

\def\mb           {{\bar{m}}}

\def\ow{{\overline{W}}}

\def\mb         {{\overline M}}

\def\ow         {{\overline W}}

\begin{document}

\begin{titlepage}
\hfill \vbox{
    \halign{#\hfil         \cr
           } 
      }  
\vspace*{20mm}
\begin{center}
{\LARGE \bf{{Dynamical supersymmetry breaking in large $N_c$ supersymmetric QCD }}}\\ 

\vspace*{15mm} \vspace*{1mm} {A. Imaanpur\footnote{ aimaanpu@modares.ac.ir} and M. Koohgard\footnote{m.koohgard@modares.ac.ir}}
\vspace*{1cm}

{\it Department of Physics, School of Sciences\\ 
Tarbiat Modares University, P.O. Box 14155-4838, Tehran, Iran}\\

\vspace*{1mm}

\vspace*{1cm}

\end{center}

\begin{abstract}
We study dynamical supersymmetry breaking in supersymmetric QCD theories for $N_f<N_c$.  We consider a model with a singlet chiral superfield coupled to the infrared meson chiral superfield through a classical superpotential. We examine the vacuum structure of this model and show that in a particular limit of the parameter space with the large $N_c$ limit, it has a vacuum that dynamically breaks supersymmetry. The supersymmetric vacuum, in this limit,  is being pushed to infinity. 
\end{abstract}

\end{titlepage}

\section{Introduction}
Supersymmetric quantum field theories have provided us with some useful toy models in which one can track the strong coupling dynamics of phenomena like quark confinement, chiral symmetry breaking, and the mass gap. Moreover, they are promising candidates to replace the standard model of particle physics. The minimal supersymmetric standard model (MSSM) is a theory which minimally embeds supersymmetry in the standard model. 

Although the MSSM can address the hierarchy problem, improve the flow of the running coupling constants so that they all meet at one high energy scale point, and provide particle candidates for dark matter, it is not yet quite clear how to consistently handle the problem of supersymmetry breaking in this model. A hierarchy of energy scales can be naturally achieved through dynamical supersymmetry breaking. In SQCD theories, for instance, this is triggered by the generation of nonperturbative superpotentials \cite{SEIO1, SEIO2, LEC}. 
In fact, depending on the number of colors, $N_c$, and the number of the flavors, $N_f$, one gets different nonperturbative superpotentials. Therefore, in principle, it is natural to think of models which classically have a supersymmetric vacuum but at the quantum level a nonperturbative superpotential is generated resulting in a nonsupersymmetric vacuum. Constructing a realistic model of this kind, however, turns out to be difficult and, in fact, it is nongeneric. For $N_c=N_f$, there are some interesting models for which F-flatness conditions are in conflict with 
the dynamical constraint on the quantum vacuum moduli space, and hence supersymmetry is dynamically broken \cite{IY, IT}. This example, however, is {\em noncalculable}, in the sense that the vacuum is not in the weakly coupled region.

The idea of having a long-lived metastable vacuum instead of a nonsupersymmetric absolute minimum opened a new avenue in constructing phenomenologically viable models of 
dynamical supersymmetry breaking \cite{SEI1, SEI2}. Metastable vacua are easier to construct and appear in more generic models. In \cite{SEI1} Seiberg {\em et al}.   proposed a SQCD model with $N_c<N_f<3/2N_c$, which, instead of having a stable nonsupersymmetric vacuum, has a metastable vacuum. They showed that, by a suitable adjustment of the parameters of the model, this vacuum can be long lived.      

In this paper, we study SQCD with $N_f<N_c$ deformed by a singlet chiral superfield. The singlets are coupled to the low energy meson chiral superfields through a classical superpotential. We examine the vacuum moduli space and observe that, apart from a supersymmetric solution, there are also nonsupersymmetric solutions. We choose a maximally symmetric solution and look at the small fluctuations around it to determine its stability. The analysis of the stability simplifies in a corner of the parameter space of the model; however, to keep the potential finite, we need to take a simultaneous large $N_c$ limit. We discuss the spectrum of small fluctuations and show that there are no tachyonic modes in this particular limit.\footnote{Large $N$ supersymmetry breaking in lower dimensions has been studied in \cite{AFF}.}  Furthermore, we show that it is also possible to keep the nonsupersymmetric vacuum in the weakly coupled region of the field space and very far from the supersymmetric one. In fact, the supersymmetric vacuum will be pushed to infinity in that limit.  

The organization of this paper is as follows. In the next section, we introduce the model and discuss its vacuum structure. We observe that the model, in addition to supersymmetric vacua, admits nonsupersymmetric extrema. To discuss the classical stability, in Sec. 3, we work out the mass matrix around a maximally symmetric solution. In Sec. 4, we look at a particular limit in the 
parameter space of the model which simplifies the stability analysis. We derive the corresponding eigenvalues and show how this vacuum can be placed far away from the supersymmetric one. Conclusions and an outlook are given in Sec. 5. In the Appendix, we derive the details of the computations of the successive derivatives of the K\"ahler potential. 

\section{The model}
Supersymmetric QCD has a dynamical mass scale $\Lambda$ below which the coupling constant becomes strong so that $N_f$ quark chiral superfields $\ph^a_i$ and $N_f$ antiquark $\tilde{\ph}^{ai}$ condense into the singlet meson chiral superfields $M_i^{\ j}=\ph_i^a \tilde{\ph}_a^j$.\footnote{Note that $\ph^a_i$ and $\tilde{\ph}^{ai}$  transform in the fundamental and antifundamental representations of the $SU(N_c)$ gauge group, respectively.} These mesons constitute the low energy degrees of freedom of the theory. Nonrenormalization theorems forbid any perturbative quantum corrections to the superpotential. However, a superpotential is generated nonperturbatively \cite{SEIO1}. For $N_f<N_c$, it reads
\be
(N_c-N_f) \left (\f{\Lambda^{3N_c-N_f}}{\det M} \right )^{\f{1}{N_c-N_f}}\, .\label{NONPER}
\ee 
With no classical superpotential, the model will only have supersymmetric runaway vacua at infinity. To modify the vacuum structure and get some possible metastable vacua, we introduce singlet (under the gauge group $SU(N_c)$) chiral superfields $S_{ij}$ 
which couple to the low energy meson fields $M_{ij}$ through the following superpotential:
\be
W_{cl}= \eta M_{ij}S^{ji} + \ga S_{ij}S^{ji} + m_{ij}M^{ji} \, ,\label{SUP}
\ee
where $\eta$, $\ga$, and $m_{ij}$ are complex parameters, with $i, j, \ldots =1, 2, \ldots, N_f$. This superpotential should be regarded as an effective superpotential, which, in principle, could be derived from a more fundamental theory. It can arise, for example, from higher dimensional operators induced from supergravity or by integrating out some massive fields at higher energies. Given the above superpotential, one can discuss the low energy effective description and especially the vacuum structure of the model. If the mass parameters of the fields $M$ and $S$ are large compared to $\Lambda$, then they can be integrated out before hitting the scale $\Lambda$. One is then left with a pure supersymmetric gauge theory which confines at $\Lambda$, and because of the mass gap, a nonperturbative superpotential is generated. Upon integrating in the $M$ fields, superpotential (\ref{NONPER}) is then obtained. On the other hand, if the mass parameters are small compared to $\Lambda$, those terms survive and need to be included in the low energy effective theory. However, because of the nonrenormalization theorems, the classical superpotential is perturbatively exact, and one only needs to add the nonperturbative superpotential (\ref{NONPER}) to the tree-level superpotential to describe the low energy effective theory \cite{LEC, LEIGH}. Hence, taking the nonperturbative superpotential into account, the superpotential for $N_f<N_c$ becomes
\be
W = (N_c-N_f) \left (\f{\Lambda^{3N_c-N_f}}{\det M} \right )^{\f{1}{N_c-N_f}} + \eta M_{ij}S^{ji} + \ga S_{ij}S^{ji} + m_{ij}M^{ji} \, .\label{WSUP}
\ee

Let us now discuss the vacuum structure of this model. First we need to compute the F-terms,
\be
\f{\pl W}{\pl M_{ij}}=  -\left (\f{\Lambda^{3N_c-N_f}}{\det M} \right )^{\f{1}{N_c-N_f}} M_{ji}^{-1} +\eta S_{ji} + m_{ji} \, ,\label{susy1}
\ee
and for $S_{ij}$ we get
\be
\f{\pl W}{\pl S_{ij}}=\et M_{ji} + 2\ga S_{ji}\, .\label{susy2}
\ee
If we set the F-terms to zero, we get the supersymmetric vacuum:
\be
-\left (\f{\Lambda^{3N_c-N_f}}{\det M} \right )^{\f{1}{N_c-N_f}} M_{ij}^{-1} -\f{\eta^2}{2\ga} 
M_{ij} + m_{ij}  =0 \, ,\label{S0}
\ee
and,
\be
S_{ij}=-\f{\et}{2\ga}M_{ij}\, .\label{S1}
\ee
For a maximally symmetric vacuum solution, we have
\be
M_{ij}= \la\, \del_{ij}\, ,\ \ \ \ \   S_{ij}={s}\, \del_{ij}\,  . \label{MAX} 
\ee
Moreover, if we set $m_{ij}=m\, \del_{ij}$, the F-term equation (\ref{S0}) for a supersymmetric solution reduces to
\be
\Lambda^{\f{3N_c-N_f}{N_c-N_f}} \la^{\f{N_c}{N_f - N_c}} + \f{\et^2}{2\ga}\la -m =0\, ,\label{SUSY}
\ee
whereas eq. (\ref{S1}) becomes
\be
{s}=-\f{\et}{2\ga}\la \, .   \label{SUSYs}
\ee 

Next, we want to examine whether there are nonsupersymmetric vacua. For the potential, with a canonical K\"ahler potential for $S_{ij}$, we have
\be
V = K^{-1}_{ij,kl}\, \f{\pl W}{\pl M_{ij}}\f{\pl \ow}{\pl \mb_{kl}}  + \lf |\f{\pl W}{\pl S_{ij}}\ri |^2 \, ,\label{POT}
\ee
where $K^{-1}_{ij,kl}$ is the inverse of the metric
\be
K_{ij,kl}= \f{\pl^2 K}{\pl M_{ij} \pl \mb_{kl}}\, ,
\ee
with $K$ the K\" ahler potential
\be
K = \tr \sm = \sm_{ii}\, .\label{KAH}
\ee
This K\"ahler potential is actually induced from the microscopic theory simply by imposing the D-flatness condition and expressing the canonical K\"ahler potential in terms of the low energy degrees of freedom $M_{ij}$ \cite{LEC}. Therefore, this form of $K$ is valid only in that region of the moduli space where $M_{ij}$ is large so that the microscopic theory is weakly coupled and the use of its canonical potential is justified. 
In the present model, we will see that the nonsupersymmetric solution can be located in the weakly coupled region of field space, and hence it makes sense to use (\ref{KAH}) as the K\"ahler potential.

For the minima we derive 
\bea
 \f{\pl V}{\pl M_{rs}}&\!\!\! =&\!\!\!  \f{\pl K^{-1}_{ij,kl}}{\pl M_{rs}} \f{\pl W}{\pl M_{ij}} \f{\pl \ow}{\pl \bar{M}_{kl}} \nn \\
&+& K^{-1}_{ij,kl} \, \lf [\hat{\Lambda}\lf( \f{1}{N_c-N_f} M_{rs}^{-1}M_{ji}^{-1} +  M_{jr}^{-1}M_{si}^{-1} \ri)\ri] \nn \\
&\!\!\! \times &\!\!\! \lf(- \bar{\hat{\Lambda}} \mb_{lk}^{-1} + \bar{\eta}\bar{S}_{lk}+\bar{ m}_{lk}  \ri) 
+\et (\bar{\et} \mb_{sr}+ 2\bar{\ga} \bar{S}_{sr}) =0\, , \label{V1}
\eea 
and
\bea
 \f{\pl V}{\pl S_{rs}}&\!\!\! =&\!\!\! \et K^{-1}_{sr,kl}\, \lf( -\hat{\Lambda}\, \mb_{kl}^{-1} + \bar{\eta}\bar{S}_{kl}+\bar{ m}_{kl}  \ri) \nn \\
&\!\!\!+& \!\!\! 2\ga  (\bar{\et} \mb_{sr}+ 2\bar{\ga } \bar{S}_{sr}) =0 \, , \label{V2}
\eea
where we have defined
\be
\hat{\Lambda} = \lf(\f{\Lambda^{3N_c-N_f}}{\det M} \right )^{\f{1}{N_c-N_f}}\, .
\ee

Now, to examine nonsupersymmetric vacua we assume that the F-terms are nonvanishing, and hence we plug (\ref{V2}) into (\ref{V1}) to obtain
\bea
&&
\lf[ \f{\pl K^{-1}_{ij,kl}}{\pl M_{rs}} \lf( -\hat{\Lambda} M_{ji}^{-1} +\eta S_{ji} + m_{ji} \, \ri) \ri. \nn \\
&+& \!\! \lf. K^{-1}_{ij,kl}\lf( \hat{\Lambda}  \f{ M_{rs}^{-1}M_{ji}^{-1}}{N_c-N_f} +  \hat{\Lambda} M_{jr}^{-1}M_{si}^{-1} 
- \f{\et^2}{2\ga}\, \del_{jr}\del_{is}\ri)\ri]\,   \f{\pl \ow}{\pl \mb_{kl}} =0  \, .  \label{V3}
\eea 
As in the case of the supersymmetric vacuum, here we can also look for a maximally symmetric extremum, where  $M_{ij}$ and $S_{ij}$ are proportional to the identity matrix as in (\ref{MAX}). In the Appendix, we show that with this assumption we have
\be
K^{-1}_{ij,kl}=4|\la|\, \del_{jk}\del_{il}\, ,\ \ \ \ \  \f{\pl K^{-1}_{ij,kl}}{\pl M_{rs}} = \f{\lab}{|\la|} \lf(\del_{jk} \del_{ir}\del_{ls} +  \del_{il} \del_{kr}\del_{sj}\ri)\, .\label{K}
\ee
Therefore, assuming supersymmetry is broken, eq. (\ref{V3}) requires that we set the first bracket in that equation to zero. With (\ref{K}), this implies
\be 
\lf(-\f{\hat{\Lambda}}{\la} +\et s + m \ri) + \f{2\hat{\Lambda}}{\xi \la} -\f{\et^2}{\ga}\, \la =0\, ,\label{SOL3}
\ee
from which we get
\be
s= \f{1}{\et}\lf(\lf(\f{\xi -2}{\xi}\ri) \f{\hat{\Lambda}}{\la} +\f{\et^2\la}{\ga}  -m\ri)\, .\label{SOL2}
\ee
Plugging $s$ into (\ref{V2}) we get an equation which determines $\la$
\be
\f{\hat{\Lambda}}{\la} \lf[\f{2}{\xi}+ \lf(\f{2}{\xi}-1\ri)\f{|\ga|^2}{|\et|^2|\la|} \ri] \nn 
-\f{\et^2}{\ga} \lf(1 +\f{3|\ga|^2}{2|\et|^2|\la|}\ri)\la +\f{m |\ga|^2}{|\et|^2|\la|} =0\, ,\label{nonssusy}
\ee
where we have defined
\be
\xi = \f{N_c-N_f}{N_c}\, ,
\ee
so that $0< \xi <1$.

\section{Classical stability}
 In this section we discuss the stability of nonsupersymmetric solutions (\ref{SOL2}) and (\ref{nonssusy}) by looking at the mass matrix of small fluctuations around them. With the canonical K\"ahler potential, the bosonic mass matrix reads
\be
\lf(
\begin{array}{ll}
 {W}_{ij}\overline{W}^{im} & W^i\overline{W}_{ijm} \\
\overline{W}^i W_{ijm} & \overline{W}_{ij} W^{im} 
\end{array}
\ri)\, , \label{MASS}
\ee
where the indices on the superpotential $W$ indicate derivatives with respect to the chiral superfields, e.g, $W_i =\f{\pl W}{\pl \ph^i}$, and so on. Since here we have $N_f^2$  chiral superfields $S_{ij}$, and $N_f^2$ chiral superfields $M_{ij}$ of  mesons in the low energy limit of SQCD, eq. (\ref{MASS}) is a $2(2N_f^2)\times 2(2N_f^2)$ mass matrix. 

To obtain the mass matrix for a generic K\"ahler potential, however, one has to directly expand the potential to quadratic order. As we have chosen a maximally symmetric vacuum, the potential of small fluctuations splits into two parts. The first part consists of quadratic fluctuations of 
symmetric (antisymmetric) matrices $\del M_{ij} \del M^{ij}$, and the second part contains the quadratic trace terms $\del M_i^{\ i} \del M_j^{\ j}$
\bea
\f{\pl^2 V}{\pl M_{kl} \, \pl  \overline{M}_{rs}}\, \del M_{kl} \del  \overline{M}_{rs} 
\!\!&=& \!\!(A\, \del _{kr}\del_{ls} + A'\, \del_{kl}\del_{rs})\, \del M_{kl} \del  \overline{M}_{rs} \nn \\
\!\!& =& \!\!A\, \del M_{kl} \del  \overline{M}^{kl} + N_f^2 A'\, \f{\del M_k^{\ k}}{N_f} 
\f{\del  \overline{M}_l^{l}}{N_f} \label{AB}\, .
\eea
Let us further decompose the symmetric part of $\del M_{ij}$ as follows:
\be
\del M_{ij} =  \f{1}{N_f} \del M_k^{\ k}\, \del _{ij} + \del \hat{M}_{ij}\, ,
\ee
where $\del M_k^{\ k}$ is the trace part, and $\del \hat{M}_{ij}$ is the traceless part. In so doing, eq. (\ref{AB}) can be decomposed into two independent degrees of freedom
\be
A\, \del M_{kl} \del  \overline{M}^{kl} + A'\, \del M_k^{\ k} \del  \overline{M}_l^{l} \label{AB}
= A\, \del \hat{M}_{kl} \del  \overline{\hat {M}}^{kl} + N_f^2(A'+ {A}/{N_f})\, \f{\del M_k^{\ k}}{N_f} 
\f{\del  \overline{M}_l^{l}}{N_f}\, .
\ee
Using the formula for the derivatives of $K^{-1}_{ijkl}$ in the Appendix to compute the second derivative of potentials (\ref{POT}) and (\ref{SOL3}), we obtain
\bea
A\!\!&=&\!\!  \f{1}{|\la|} \lf|2\lf(\xi -1 \ri)\f{\hat{\Lambda}}{\xi\la}  +\f{\et^2}{\ga}\la \ri|^2   +  |\et|^2 \label{A} \\
A_t &\equiv & N_f^2\lf(A' + \f{A}{N_f}\ri) = {N_f}|\la|\lf|\f{\et^2}{\ga }\ri|^2 + N_f|\et |^2  \, .\label{At}
\eea
Similarly, for $\del S_{kl} \del \overline{S}_{rs}$ we get
\bea
\f{\pl^2 V}{\pl S_{kl} \pl \overline{S}_{rs}}\, \del S_{kl} \del  \overline{S}_{rs} 
 \!\!&=&\!\! (B\, \del _{kr}\del_{ls})\, \del S_{kl} \del  \overline{S}_{rs}  \nn \\
 \!\!&=&\!\! B\, \del \hat{S}_{kl} \del  \overline{\hat {S}}^{kl} + 
B_t\, \f{\del S_k^{\ k}}{N_f} \f{\del  \overline{S}_l^{l}}{N_f}  \label{}\, ,
\eea
where
\be 
B = 4 |\la| |\et|^2 + 4|\ga |^2  \, ,\label{B}
\ee
and
\be
B_t=N_f B  \, .\label{Bt}
\ee

Next, consider quadratic fluctuations of $\del S_{kl} \del \overline{M}_{rs}$,
\bea
\f{\pl^2 V}{\pl S_{kl} \pl \overline{M}_{rs}}\, \del S_{kl} \del  \overline{M}_{rs} 
\!\!&=&\!\! (\al\, \del _{kr}\del_{ls} +\al'\, \del_{kl}\del_{rs})\, \del S_{kl} \del  \overline{M}_{rs} \nn \\
\!\!&=&\!\! \al\, \del \hat{S}_{kl} \del  \overline{\hat {M}}^{kl} + N_f^2(\al '+{\al}/{N_f})\, \f{\del S_k^{\ k}}{N_f} \f{\del  \overline{M}_l^{l}}{N_f}   \label{}\, ,
\eea
where
\bea
&&\al = 2|\la| \bar{\et} \lf(2\lf(1-\f{1}{\xi}\ri)\f{\hat{\Lambda}}{\la^2}+\f{\et^2}{\ga} \ri) +2\et\bar{\ga}   \label{AL} \\
&&\al_t=N_f^2\lf(\al ' +\al/N_f\ri) = \f{ {2\et N_f}}{{\ga} } ( |\la| |\et|^2 + |\ga |^2) 
=\f{{\et}}{{2\ga} } B_t\, .\label{ALt}
\eea

Finally, we compute the off-diagonal elements in (\ref{MASS}). For $\del S_{kl} \del {M}_{rs}$ we get
\bea
&&\f{\pl^2 V}{\pl S_{kl} \pl {M}_{rs}}\, \del S_{kl} \del  {M}_{rs} = (e\, \del_{kl}\del_{rs}) \del S_{kl} \del {M}_{rs}\, ,
\eea
where
\be
e = 2  \lf(-\f{2\hat{\Lambda}}{\xi \la^2} +\f{{\et}^2}{{\ga}} \ri)
\!|\la|\et \, ,
\ee
with
\be
e_t = N_f\, e \, .\label{et}
\ee
Moreover, the coefficient of  $\del M_{kl} \del {M}_{rs}$ reads
\bea
 \f{\pl^2 V}{\pl M_{kl} \pl {M}_{rs}}\, \del M_{kl} \del {M}_{rs} 
\!\!\!&=&\!\!\! (C\, \del _{kr}\del_{ls} +C'\, \del_{kl}\del_{rs})\, \del M_{kl} \del  {M}_{rs} \nn \\
\!\!\!&=&\!\!\! C\, \del \hat{M}_{kl} \del  {\hat {M}}^{kl} + N_f^2(C '+{C}/{N_f})\, \f{\del M_k^{\ k}}{N_f} \f{\del {M}_l^{l}}{N_f} \, ,  \nn
\eea
where
\be
C=4{|\la|}  \lf(\f{2\bar{\hat{\Lambda}}}{\xi\bar{\la}^2} - \f{\bar{\et}^2}{\bar{\ga}} \ri) \!\! \lf(\f{\hat{\Lambda}}{2\xi \la^2} +\f{\et^2}{4\ga}\ri)\!\! \lf(\f{\bar{\la}}{\la}\ri)\, , \label{C}
\ee
and
\bea
C_t \!\! &=&\!\! N_f^2\lf( C' + C/N_f\ri)  \nn \\
\!\! &=&\!\! 4N_f{|\la|}  \lf(\f{2\bar{\hat{\Lambda}}}{\xi\bar{\la}^2}  -\f{\bar{\et}^2}{\bar{\ga}} \ri) \!\!\lf( \lf(2-\xi\ri)\f{\hat{\Lambda}}{2\xi^2\la^2} +\f{\et^2}{4\ga} \ri)\!\! \lf(\f{\bar{\la}}{\la}\ri)\, \label{Ct}
\eea

We are now ready to write down the mass matrix (\ref{MASS}). For the traceless part it reads
\be
\lf(
\begin{array}{llll}
 A & \al & C & e \\
\bar{\al}  & B & e & 0 \\
\bar{C} & \bar{e} & A & \bar{\al} \\
\bar{e} & 0 & {\al} & B 
\end{array}
\ri)\, , \label{MAT}
\ee
where $A, B, \al $, and $C$ are derived in (\ref{A}), (\ref{B}), (\ref{AL}), and (\ref{C}), respectively. There is a similar matrix 
for the trace part with components $A_t, B_t, \al_t $ and $C_t$ as defined in (\ref{At}), (\ref{Bt}), (\ref{ALt}), and (\ref{Ct}), respectively. 

Let us discuss the trace part, where we can identify the Goldstino.
In diagonalizing (\ref{MAT}), we obtain the following characteristic equation of the eigenvalues
\be
\lf(|e_t|^2 +|\al_t|^2 - A_t B_t +k_t(A_t +B_t) -k_t^2\ri)^2 =\lf| (B_t - k_t) C_t -2e_t\al_t \ri|^2\, .\label{K2}
\ee
For the mass matrix of fermions, on the other hand, the off-diagonal $2\times 2$ matrices of (\ref{MAT}) are zero. Moreover, from (\ref{At}), (\ref{Bt}), and (\ref{ALt}) we deduce 
\be
A_t B_t = |\al_t|^2\, ,
\ee
and so, using this in  (\ref{K2}), we see that for fermions
\be
 k_t^2-k_t(A_t +B_t) =0\, .
 \ee
Hence, we get a massless mode which we identify as the {\em Goldstino} coming from the supersymmetry breaking on 
solution (\ref{nonssusy}).

\section{Large $N_c$ limit}
There are certain limits in which eq. (\ref{K2}) reduces to a quadratic equation and the analysis of the stability simplifies. 
For instance, we can take either $C_t$ or $B_t C_t -2e_t\al_t$ in (\ref{K2}) to vanish. Together with (\ref{nonssusy}), this constrains the space of parameters of the model. However, in both of these cases we will end up with some negative eigenvalues of (\ref{K2}) resulting in the instability of the solution. 

Alternatively, we can go to a limit where both $\et$ and $\ga$ vanish so that eq. (\ref{nonssusy}) 
and the stability analysis greatly simplify. However, a look at the potential shows that to have a finite value for $V$, we need to take a large $N_f$ limit simultaneously. Thus, we take the following limit
\be
\et = C_1/\sqrt{N_f}\, , \ \ \ \ \ \ga  = C_2/N_f \, ,\label{LIMIT0}
\ee
with $C_1$ and $C_2$ two finite complex constants (with $\et^2/\ga$ fixed) and large $N_f$. Explicitly, first note that from eqs. (\ref{V2}) and (\ref{SOL3}) we see that the extremum of the potential satisfies the following equations: 
\be 
-\f{\hat{\Lambda}}{\la} +\et s + m  = -\f{\bar{\ga}}{2|\la|\bar{\et}}\lf(\et \la +2\ga s\ri)
=- \f{2\hat{\Lambda}}{\xi \la} + \f{\et^2}{\ga}\la  \, .\label{ROOT}
\ee
So assuming that $\la$ and $s$ at the extremum are finite, the last two equations of (\ref{ROOT}) imply that in the limit of (\ref{LIMIT0}) we have
\be
- \f{2\hat{\Lambda}}{\xi \la} + \f{\et^2}{\ga}\la = - \f{C_2 \la}{2N_f |\la|}\lf(\f{\et}{\bar{\et}}\ri) + {\cO}(1/N_f^{3/2})\, ,\label{ORDER}
\ee
where we have used (\ref{LIMIT0}) to replace for $\ga$ on the right-hand side. Now we can use (\ref{LIMIT0}) and (\ref{ROOT}) to see how  potential (\ref{POT}) scales at large $N_f$,
\bea
V&=&4N_f|\la| \lf|-\f{\hat{\Lambda}}{\la} +\et s + m \ri|^2 + N_f\lf|\et \la +2\ga s\ri|^2 \nn \\
&=& \lf(N_f \lf| \f{\ga}{\et} \ri|^2 \f{1}{|\la|} + N_f\ri) \lf|\et \la +2\ga s\ri|^2 \nn \\
&=& 4|\la|^2 \lf( \lf| \f{C_2}{C_1} \ri|^2 \f{1}{|\la|} + N_f\ri) \lf| \f{\et}{\ga} \ri|^2  \lf| - \f{2\hat{\Lambda}}{\xi \la} + \f{\et^2}{\ga}\la \ri|^2 \nn \\
&= &  |C_1 \la|^2 + \cO(1/N_f)\, , \label{POT3}
\eea
where in the last equality we have used (\ref{ORDER}). Hence, we observe that in the limit (\ref{LIMIT0}) the potential, evaluated on the extremum, approaches a finite value.

To discuss the mass matrix in the limit (\ref{LIMIT0}), first note that, assuming $\la$ is finite, from (\ref{ROOT}) we infer
\be
\lf(-\f{2\hat{\Lambda}}{\xi\la^2} +\f{\et^2}{\ga}\ri) \sim 1/{N_f}\, .\label{SCALE}
\ee
Now, looking at the elements of the trace part of the mass matrix, $A_t, B_t, \al_t, e_t $, and $C_t$ -- as derived in (\ref{At}), (\ref{Bt}), (\ref{ALt}), (\ref{et}), and (\ref{Ct}) --  eq. (\ref{SCALE}) implies that, in the large $N_f$ limit they scale as 
\be
N_f,\, 1,\,  N_f^{1/2}\, ,  N_f^{-1/2}\, , 1 \, ,
\ee
respectively. We can now analyze the quartic characteristic eigenvalue equation (\ref{K2}). Using the explicit expressions for the roots of this equation and keeping only the dominant terms in the large $N_f$ limit, we find that the eigenvalues scale as follows:\footnote{We have also  checked the leading order terms using Mathematica.}
\be
{\rm diag }\lf(N_f,\,  N_f,\, 1/N_f,\,  -1/N_f \ri) \, .\label{EIG1}
\ee
Hence, in the limit $N_f\to \infty$, the first  two modes get very heavy and decouple from the spectrum, and we are left with two zero-mode eigenvalues. For the nontrace part, the matrix elements, $A, B, \al, e $, and $C$, in the large $N_f$ limit scale as
\be
1,\,   {N_f}^{-1},\,   N_f^{-1/2}, \,   {N_f}^{-3/2},\,  N_f^{-1}\, ,
\ee
respectively. Analyzing the quartic eigenvalue equation, we find that the leading terms of the eigenvalues scale as follows:
\be
{\rm diag }\lf( 1,\,  1,\, 1/N_f ,\, -1/N_f \ri) \, ,\label{EIG2}
\ee
so in the limit  $N_f\to \infty$ we get two extra zero modes. Note that there is a similar expression for the modes coming from the antisymmetric part of the quadratic fluctuations, so, in sum, we get 
$2 N_f^2$ real massless scalars. Here we get a factor of $2$ as matrix (\ref{MAT}) is $4\times 4$, and hence the eigenvalues are doubly degenerate. We conclude that in the large $N_f$ limit, there are no negative eigenvalues of the mass matrix of small fluctuations around the nonsupersymmetric solution, and hence no instability. 

Now let us discuss what happens to solution (\ref{nonssusy}) in the limit $\ga/\et \to 0$ and large $N_c$. Setting  
$N_c=aN_f$, with $a>1$, and since $\et^2/\ga$ is held fixed and finite, solution (\ref{nonssusy}) in this limit reduces to the following equation for a {\em nonsupersymmetric} vacuum:
\be
\lf(\f{\Lambda^{{3N_c-N_f}} }{\la^{{2N_c-N_f}}}\ri)^{\f{1}{N_c-N_f}} = \f{\et^2}{2\ga } \lf(\f{N_c-N_f}{N_c}\ri)  \, . \label{SOL1}
\ee
Note that this equation is the same as (\ref{ORDER}) in the large $N_f$ limit. Further, let us define
\be
\f{\et^2}{2\ga } \lf(\f{N_c-N_f}{N_c}\ri) \equiv \f{1}{\Lambda L^{\f{2N_c-N_f}{N_c-N_f}}}\, ,\label{LIMIT}
\ee
where $L > 1$ is a finite number, and on the right-hand side we have included a factor of $\Lambda$ for dimensions to match. Equation ({\ref{SOL1}) now implies
\be
\la =  L \Lambda^2\, . \label{RL}
\ee
Therefore, we observe that, by taking $L\gg 1$, the nonsupersymmetric solution (\ref{RL}) can be located in the weakly coupled (in terms of the microscopic degrees of freedom) region of the field space. As mentioned in Sec. 2, this also justifies the use of the K\"ahler potential (\ref{KAH}) induced from the microscopic theory. Further, note that eq. (\ref{ROOT}) shows that, in the limit of vanishing $\et$ and $\ga$, for any finite and constant $s$ we have a solution. Therefore, $s$ is a modulus, and this coincides exactly with $2N_f^2$ zero modes in the large $N_f$ limit that we obtained in (\ref{EIG1}) and (\ref{EIG2}). 

Having obtained $\la$, we can now derive the explicit minimum value of the potential in the large $N_f$ limit. From (\ref{POT3}) and (\ref{RL}), we have
\be
V= |C_1|^2 |\la|^2 = L^2|C_1|^2 |\Lambda|^4\, ,\label{POTT}
\ee
which gives an energy scale for the nonsupersymmetric small excitations in this vacuum, and hence a scale of supersymmetry breaking:
\be
M_s^4 = L^2|C_1|^2 |\Lambda|^4\, .\label{MS}
\ee
So, if we take 
\be
L|C_1| \ll 1\, , 
\ee
then $M_s\ll \Lambda$, and supersymmetry breaking happens at a scale well below $\Lambda$ which, in turn, 
justifies the use of meson low energy degrees of freedom to describe this region of field space.  

In the end, let us examine the supersymmetric vacuum and see how it compares with the nonsupersymmetric one in the limit. The supersymmetric solutions (\ref{susy1}) and (\ref{susy2}) are
\bea
&& -\f{\hat{\Lambda}}{\la} +\et s + m =0\, , \nn \\
&&\ \   \et \la + 2\ga {s}=0 \, . 
\eea
The second equation implies $\et s= -\f{\et^2}{2\ga}\la$ and since we have kept $\et^2/\ga$ fixed, $\et s$ is finite. Plugging $\et s$ into the first equation, we can derive $\la$. However, since $s=-\f{\et}{2\ga}\la$, it diverges in the limit (\ref{LIMIT0}). 
Therefore, for the supersymmetric solution, $s$ is located at infinity. On the other hand, recall that the model also admits nonsupersymmetric solutions with finite $s$. The moduli space of nonsupersymmetric solutions is thus the whole complex plane with the potential at its constant value (\ref{POTT}) approaching zero at the boundary ($s \sim \sqrt{N_f}$; the supersymmetric solution). 
Therefore, any nonsupersymmetric solution with finite $s$ is very far from the supersymmetric solution, and since the potential along the modulus is flat it is also stable. This can also be seen from the Coleman bounce action \cite{COL}
\be
S_B \sim \f{(\Delta M)^2}{V_{meta}}\, ,
\ee 
where $\Delta M$ is the distance between the metastable and supersymmetric vacua and $V_{meta}$ is the metastable potential. In the large $N_c$ limit, $V_{meta}$ is finite, whereas $\Delta M$ goes to infinity; hence the probability of quantum tunneling approaches zero.

\section{Conclusions}
Dynamical supersymmetry breaking in models with metastable vacua have so far been extensively studied. These models, in particular, serve as the hidden sector of 
MSSM-like theories where supersymmetry first breaks and then is mediated to the visible sector. In SQCD, for instance, interesting generic models for $N_c=N_f$ and $N_c<N_f<3/2N_c$
have been constructed  that admit metastable vacua \cite{IY, IT, SEI1}. In this paper, we addressed the existence of such  vacua  for $N_f<N_c$. 

We showed that supersymmetric QCD with $N_f<N_c$ coupled to a singlet chiral superfield can have stable nonsupersymmetric vacua. We observed that apart from a supersymmetric solution, the model also admits nonsupersymmetric solutions. We 
discussed the mass matrix of small fluctuations around a maximally symmetric solution.  
We noticed that there are certain limits on the space of parameters where the quartic equation of eigenvalues reduces to a quadratic equation, and thus the question of stability is tractable. However, the appearance of  tachyonic modes in such cases led us to examine the potential and consider instead a limit of large $N_f$ with $N_f \ga^2/\et^2$ kept fixed. We analyzed the mass matrix of small fluctuations and showed that the spectrum, in this limit, contains no tachyons. 

In the large $N_f$ limit, the spectrum showed some extra zero modes. We argued that these zero modes correspond to the 
pseudomoduli space of flat directions along $s$. The expectation value of the meson field on the field space turned out to be proportional to a finite number $L$, which we took to be large for the solution to be in  
the weakly coupled region of field space. Further, as the supersymmetric solution gets pushed to infinity in the limit, the 
two solutions become infinitely far apart. The nonsupersymmetric vacuum thus remains stable under small fluctuations. 

It is very interesting to further examine the characteristic equation and see if there are any other limits where the spectrum of the mass matrix is free from tachyons. In this model, we encountered a pseudomoduli space of flat directions along $s$ which was the characteristic of the limit taken in the parameter space of the model. It is important to see whether this modulus is lifted when quantum corrections are included. 

\vspace{1 cm}

\appendix
\section{Computation of the successive derivatives of the K\"ahler potential}
Although computing the derivative of the square root of a matrix in terms of the derivative of the matrix itself is not straightforward, in this appendix, we show how to derive such derivatives calculated on the solutions of type $M_{ij}=\la \del_{ij}$. This is done by taking the successive derivatives. In each step, we compute the value of the derivative on the solution, and then use it to compute the derivative in the next step. Let us start from the definition of the square root of a matrix,
\be
\mb_{il} M_{lj} = \sm_{il}\sm_{lj}\, .
\ee
Taking the first derivative  we deduce
\be
M_{sj}\del_{ir} = \f{\pl \sm _{il}}{\pl \mb_{rs}}\sm_{lj} +\sm_{il} \f{\pl \sm_{lj}}{\pl \mb_{rs}}\, . \label{DED}
\ee
Multiplying from right by $\sm^{-1}$ results in 
\be
M_{sj}\del_{ir}\sm^{-1}_{jk} = \f{\pl \sm_{ik}}{\pl \mb_{rs}} +\sm_{il} \f{\pl \sm_{lj}}{\pl \mb_{rs}}\sm^{-1}_{jk}\, .
\ee
Now, summing over $i$ and $k$,
\be
2\, \f{\pl \sm_{ii}}{\pl \mb_{rs}} = M_{sj}\sm^{-1}_{jr} = \mb^{-1}_{sj}\sm_{jr}\, ,
\ee
and then taking the second derivative gives 
\be
2\, \f{\pl^2 \sm_{ii}}{\pl M_{pq}\, \pl \mb_{rs}} = \mb^{-1}_{sj}\f{\pl\sm_{jr}}{\pl M_{pq}}\, .\label{SECOND0}
\ee

Since, in the end, we need the value of the derivatives on the solution, let us set $M_{ij}= \la \del_{ij}$ and $\sm_{ij} = |\la| \del_{ij}$ in (\ref{DED}) to obtain 
\be
 \f{\pl \sm_{ij}}{\pl \mb_{rs}} =\f{\la}{2|\la|}\, \del_{ir} \del_{js}\, . \label{FIRST}
\ee
Using this in (\ref{SECOND0}) we get
\be
K_{pq,rs}\equiv \f{\pl^2 K}{\pl M_{pq}\, \pl \mb_{rs}} =\f{\pl^2 \sm_{ii}}{\pl M_{pq}\, \pl \mb_{rs}} = \f{\la^2}{4|\la|^3}\, \del_{sp}\del_{rq}
\ee
on the solution.

Taking the second derivative of (\ref{DED}) we find
\bea
\del_{sp}\del_{jq}\del_{ir} &=& \f{\pl^2 \sm _{il}}{\pl M_{pq}\pl \mb_{rs}}\sm_{lj} +\f{\pl \sm_{il}}{\pl M_{pq}} \f{\pl \sm_{lj}}{\pl \mb_{rs}}\nn \\
&+&\f{\pl \sm_{il}}{\pl \mb_{rs}} \f{\pl \sm_{lj}}{\pl M_{pq}} +\sm_{il} \f{\pl^2 \sm _{lj}}{\pl M_{pq}\pl \mb_{rs}}\, , \label{DED2}
\eea
which, upon using (\ref{FIRST}), on the solution reads
\be
\f{\pl^2 \sm _{ij}}{\pl M_{pq}\pl \mb_{rs}}=\f{1}{8|\la|}\lf(3\del_{sp}\del_{jq}\del_{ir}
- \del_{ip}\del_{rq}\del_{js} \ri)\, . \label{SECOND}
\ee 
Similarly
\be
\f{\pl^2 \sm _{ij}}{\pl M_{pq}\pl M_{rs}}=-\f{\bar{\la}^2}{8|\la|^3}\lf(\del_{sp}\del_{jq}\del_{ir}
+ \del_{ip}\del_{rq}\del_{js} \ri)\, .\label{SECOND1}
\ee 

Next, we take the derivative of (\ref{DED2}) and then use the value of the first and the second derivatives on the solutions (\ref{FIRST}), (\ref{SECOND}), and (\ref{SECOND1}) to get
\be
\f{\pl K_{pq,rs}}{\pl \mb_{mn}}=\f{\pl^3 \sm _{ii}}{\pl \mb_{mn}\pl M_{pq}\pl \mb_{rs} }=\f{-1}{16\bar{\la}|\la|}\lf(\del_{sm}\del_{pn}\del_{rq}
+ \del_{sp}\del_{mq}\del_{rn} \ri)\, ,
\ee 
and
\be
\f{\pl K_{pq,rs}}{\pl M_{mn}}=\f{\pl^3 \sm _{ii}}{\pl M_{mn}\pl M_{pq}\pl \mb_{rs}}=\f{-\bar{\la}}{16|\la|^3}\lf(\del_{ps}\del_{mq}\del_{rn}
+ \del_{qr}\del_{ms}\del_{np} \ri)\, .\label{FD}
\ee 

By iteration of the above procedure, we obtain the fourth derivatives, 
\bea
\f{\pl^2 K_{pq,rs}}{\pl M_{ef}\pl \mb_{mn}}=\f{\pl^4 \sm _{ii}}{\pl M_{ef}\pl \mb_{mn}\pl M_{pq}\pl \mb_{rs}}\!\!& =& \!\! \f{-1}{32|\la|^3}
\lf( -\del_{sm}\del_{pn}\del_{eq}\del_{rf} + \del_{qm}\del_{en}\del_{sp}\del_{rf}  \ri. \nn \\
\!\! &-&\!\! \lf. \del_{sm}\del_{en}\del_{pf}\del_{rq}  + \del_{fm}\del_{pn}\del_{rq}\del_{se} \ri. \nn \\
\!\!&-&\!\! \lf. \del_{fp}\del_{mq}\del_{rn}\del_{se} - \del_{sp}\del_{qe}\del_{mf}\del_{rn} \label{FORTH}
\ri)\, ,
\eea 
with
\bea
\f{\pl^2 K_{pq,rs}}{\pl \mb_{ef}\pl \mb_{mn}}=\f{\pl^4 \sm _{ii}}{\pl \mb_{ef}\pl \mb_{mn}\pl M_{pq}\pl \mb_{rs}}\!\!& = &\!\! \f{-1}{32\bar{\la}^2|\la|}
\lf( 3\del_{se}\del_{fm}\del_{pn}\del_{rq} + \del_{se}\del_{fp}\del_{mq}\del_{rn}  \ri. \nn \\
\!\! &+&\!\! \lf. 3\del_{sm}\del_{en}\del_{pf}\del_{rq}  + \del_{sm}\del_{pn}\del_{eq}\del_{rf} \ri. \nn \\
\!\!&-&\!\! \lf. \del_{sp}\del_{eq}\del_{mf}\del_{rn} - \del_{sp}\del_{qm}\del_{en}\del_{rf} 
\ri)\, . \label{FORTH2}
\eea 
Finally, we obtain
\be
\f{\pl K_{ij,kl}^{-1}}{\pl M_{mn}} = -  K_{ij, pq}^{-1}  \f{\pl K_{pq,rs}}{\pl M_{mn}}K_{rs, kl}^{-1}\, ,\label{K22}
\ee
where the inverse of the metric on the solution reads
\be
K_{pq,rs}^{-1} = 4|\la| \del_{qr}\del_{ps}\, .
\ee
Eq. (\ref{K22}), together with (\ref{FD}), yields
\be
\f{\pl K^{-1}_{ij,kl}}{\pl M_{mn}} = \f{\lab}{|\la|} \lf(\del_{jk} \del_{im}\del_{ln} +  \del_{il} \del_{km}\del_{nj}\ri) \, ,\label{K3}
\ee
as claimed in the text. Taking the next derivative of (\ref{K22}), and using (\ref{FORTH}), (\ref{FORTH2}), and (\ref{K3}), we derive the second derivative of the inverse metric on the solution which we needed to calculate the mass spectrum.

\hspace{30mm}


\end{document}